\documentstyle[12pt,a4,psfig]{article}
\def\xa{x_\alpha^{}}
\def\xibar{\bar{\xi}}
\def\to{\rightarrow}

\def\zzp{$Z$-$Z'$\,}

\def\vsp#1{\vspace{#1 cm}}
\def\ov{\overline}

\def\lsim{{\mathop <\limits_\sim}}

\def\gm5{\gamma_5}

\def\mt{m_t^{}}
\def\mh{m_H^{}}
\def\xt{x_t^{}}
\def\xh{x_H^{}}
\def\mh{m_H^{}}

\def\sbar{\bar{s}^2}

\def\gzbar{\bar{g}_Z^2}
\def\gwbar{\bar{g}_W^2}
\def\abar{\bar{\alpha}}

\def\etal{{\it et al.}}
\newcommand{\beq}{\begin{equation}}
\newcommand{\eeq}{\end{equation}}
\newcommand{\bea}{\begin{eqnarray}}
\newcommand{\eea}{\end{eqnarray}}
\newcommand{\bsub}{\begin{subequations}}
\newcommand{\esub}{\end{subequations}}

\renewcommand{\theequation}{\thesection.\arabic{equation}}
\newcommand{\clean}{\setcounter{equation}{0}}

\def\lsim{\raisebox{0.6mm}{$<$}\hspace{-3.2mm}\raisebox{-1.3mm}{$\sim$}}
\def\extra_charge{
	\begin{table}[t]
	\begin{center}
	\begin{tabular}{|l|r|r|r|r|}
	\hline
	        & $\delta$ & $Q_E^Q$ &  $Q_E^U$ & $Q_E^D$ \\ \hline
	model A~\cite{eta_model} & 1/3
	 & $-5/18$ & $+10/18$ & $-5/18$ \\ \hline
	model B~\cite{flipped}  & 0
	& $+5/18$ &   0      & $-5/18$ \\ \hline
	model C~\cite{lbl}      & 0
	& $-5/18$ & $+10/18$ & $+5/18$ \\ \hline
	\end{tabular}
	\caption{{\small ${\rm U(1)'}$ charge assignments on $u$ 
		and $d$ quarks in the three leptophobic 
		models~\cite{eta_model,flipped,lbl}. 
		The effective ${\rm U(1)'}$ charge $Q_E$ is defined 
		as $Q_E \equiv Q'_E + \delta Y$. 
		The kinetic mixing parameter $\delta$ is taken to 
		be 1/3 for the model A~\cite{eta_model} and 0 for the 
		other two models~\cite{flipped,lbl} in 
		order to make these models leptophobic. 
		}}
	\label{tab:charge}
	\end{center}
	\vsp{0.3}
	\end{table}
	}
\def\lep_table{
	\begin{table}[tp]
	\begin{center}
	\begin{tabular}{|r|c|r|r|r|r|}
	\hline
	\multicolumn{2}{|c|}{} & 
	\multicolumn{4}{c|}{pull $\; = \;  
	 \frac{\langle {\rm data}
	 \rangle -{\rm prediction}}{
	\langle {\rm error} \rangle}$} \\ \cline{3-6}
	\multicolumn{2}{|c|}{} &  SM & A & B & C  \\ \hline
	$Z\;\;\;\;\;\;\;\;\;\;m^{}_Z$ (GeV) & 91.1867$\pm\;$0.0020&&&&\\ 
	$\Gamma_Z^{}$ (GeV)  & 2.4948  $\pm$ 0.0025  
	& $-$0.7 & $-$0.7 & $-$0.9 & $-$1.0 \\
	$\sigma^0_h$(nb) & 41.486 $\pm$ 0.053$\;\:$  
	& 0.3 & 0.3 & 0.4 & 0.4 \\ 
	$R_{\ell}$ & 20.775 $\pm $ 0.027$\;\:$ 
	& 0.9 & 1.0 & 0.7 & 0.7 \\
	$A^{0,\ell}_{FB}$ & 0.0171 $\pm $ 0.0010
	& 0.8 & 0.8 & 0.8 & 0.8 \\
	$A_{\tau}$& 0.1411 $\pm$ 0.0064 
	& $-$1.0 & $-$1.0 & $-$1.0 & $-$1.0 \\
	$A_{e} $  & 0.1399 $\pm$ 0.0073 
	& $-$1.0 & $-$1.1 & $-$1.0 & $-$1.0 \\
	$R_b$ & 0.2170 $\pm$ 0.0009 
	& 1.4 & 0.9 & 1.1 & $-$0.2 \\
	$R_c$ & 0.1734 $\pm$ 0.0048 
	& 0.3 & 0.4 & 0.3 & 0.7 \\
	$A^{0,b}_{FB}$ & 0.0984 $\pm$ 0.0024 
	& $-$2.1 & $-$2.2 & $-$2.1 & $-$2.0\\
	$A^{0,c}_{FB}$ & 0.0741 $\pm$ 0.0048 
	& 0.0 & 0.0 & 0.1 & $-$0.1 \\
	$A^0_{LR}$ & 0.1547 $\pm$ 0.0032 
	& 2.2 & 2.2 & 2.3 & 2.2 \\
	$A_b$ & 0.900 $\pm$ 0.050 
	& $-$0.7 & $-$0.7 & $-$0.7 & $-$0.7 \\
	$A_c$ & 0.650 $\pm$ 0.058 
	& $-$0.3 & $-$0.4 & $-$0.3 & $-$0.4 \\ \hline
	$W\;\;\;\;\;\;\;m^{}_W$ (GeV)&80.43 $\pm$ 0.084
	&0.6 &0.6 &0.6 &0.5 \\ \hline
	LENC$\;\;\;\;\;\;\;\;\;
	A_{\rm SLAC}$ & $\;$0.80 $\;\:\pm$ 0.058 
	& 1.0 & 0.9 & 1.0 & 0.9 \\
	$A_{\rm CERN}$ & $-$1.57 $\;\:\pm$ 0.38 $\;\;\:$
	& $-$0.4& $-$0.4 & $-$0.4& $-$0.4 \\
	$A_{\rm Bates}$ & $-$0.137 $\pm$ 0.033 $\;$
	& 0.5 & 0.5 & 0.5 & 0.4 \\
	$A_{\rm Mainz}$ & $-$0.94 $\;\,\pm$ 0.19 $\;\;\;$
	& $-$0.3 & $-$0.3 & $-$0.3 & $-$0.3 \\
	$Q_W(^{133}_{55}Cs )$ & $-$72.08 $\pm$ 0.92$\;\;\;\;\,$ 
	& 1.0 & 1.3& 1.2 & 0.5 \\
	$K_{\rm FH}\;\;\;\,$ &  0.3247 $\pm$ 0.0040 
	& $-$1.5 & $-$1.4 & $-$1.5 & $-$1.5 \\
	$K_{\rm CCFR}$ &  0.5820 $\pm$ 0.0049 
	& $-$0.5 & $-$0.5 & $-$0.6 & $-$0.5 \\ \hline 
	$\chi^2_{\rm tot}$ & & 21.8 & 21.5 & 21.6 & 18.9\\ 
	d.o.f. & & 21 & 19 & 19 & 19 \\ \hline \hline
	best fit $\;$ 
	$m^{}_t$(GeV) & & 171.6 & 172.0 & 171.6 &172.6\\
	$\alpha^{}_s (m^{}_{Z_1})$ & & 
	0.1185 & 0.1189 & 0.1176 & 0.1176\\
	$1/\bar{\alpha} (m^2_{Z_1})$ & &
	128.75 & 128.75 & 128.75 &128.74\\
	$T_{\rm new}$ & & --- & 0 & 0 & 0\\
	$\bar{\xi}$   & & --- & 0.0016 & $-$0.0010 & 0.0047\\
	\hline
	\end{tabular}
	\caption{{\small Electroweak measurements for the  
	$Z$-boson parameters, the $W$-boson mass and 
	the low-energy neutral current(LENC) experiments. 
	The best-fits and the 
	corresponding `pull' factors for the SM, the models A, 
	B and C are obtained for $m^{}_H$ = 100 GeV under the 
	constraints $m^{}_t$ = 175.6 $\pm$ 5.5 GeV~\cite{cdfd0}, 
	$\alpha^{}_s$($m^{}_{Z_1}$) = 0.118 $\pm$ 0.003~\cite{pdg96},
	1/$\bar{\alpha}(m^2_{Z_1})$ = 128.75 $\pm$
	0.09~\cite{ej95} and $T_{\rm new} \ge 0 $. 
 	Correlation matrix elements of the $Z$ line-shape parameters
	and those for the heavy-quark parameters are found in 
	ref.~\cite{lep_slc_97}.}}
	\end{center}
	\label{tab:lep_table}
	\end{table}
	}
\def\cvca_tab{	
	\begin{table}[t]
	\begin{center}
	\begin{tabular}{|l|c|c|}
	\hline
	& ${\it C_{fV}}$ & ${\it C_{fA}}$  \\ \hline
	$u$     & 3.1166 + 0.0030$x_s$ & 3.1351 + 0.0040$x_s$ \\ \hline
	$d = s$ & 3.1166 + 0.0030$x_s$ & 3.0981 + 0.0021$x_s$ \\ \hline
	$c$     & 3.1167 + 0.0030$x_s$ & 3.1343 + 0.0041$x_s$ \\ \hline
	$b$     & 3.1185 + 0.0030$x_s$ & 3.0776 + 0.0030$x_s$ \\ \hline
        $\nu$   & 1 & 1 \\ \hline
	$e=\mu$ & 1 & 1 \\ \hline
	$\tau$  & 1 & 0.9977 \\ \hline
	\end{tabular}
	\end{center}
	\caption{{\small Numerical values of factors 
	$C_{fV}, C_{fA}$ for quarks and leptons. 
	The finite mass corrections and the final state QCD 
	corrections for quarks are taken into accounted. 
	The $\alpha_s$-dependence of the QCD correction is 
	parametrized by 
	$x^{}_s = (\alpha_s(m^{}_{Z_1})-0.118)/0.003$.}}
	\vsp{0.3}
	\label{tab:cvca}
	\end{table}
	}
\def\PRD#1#2#3{Phys. Rev. {\bf D#1} (19#2) #3}
\def\NPB#1#2#3{Nucl. Phys. {\bf B#1} (19#2) #3}
\def\ZPC#1#2#3{Z. Phys. {\bf C#1} (19#2) #3}
\def\PLB#1#2#3{Phys. Lett. {\bf B#1} (19#2) #3}
\def\PRL#1#2#3{Phys. Rev. Lett. {\bf #1} (19#2) #3}
\def\PR#1#2#3{Phys. Rep. {\bf #1} (19#2) #3}
\makeatletter
\newtoks\@stequation

\def\subequations{\refstepcounter{equation}%
  \edef\@savedequation{\the\c@equation}%
  \@stequation=\expandafter{\theequation}
  \edef\@savedtheequation{\the\@stequation}
  \edef\oldtheequation{\theequation}%
  \setcounter{equation}{0}%
  \def\theequation{\oldtheequation\alph{equation}}}

\def\endsubequations{%
  \ifnum\c@equation < 2 \@warning{Only \the\c@equation\space subequation
    used in equation \@savedequation}\fi
  \setcounter{equation}{\@savedequation}%
  \@stequation=\expandafter{\@savedtheequation}%
  \edef\theequation{\the\@stequation}%
  \global\@ignoretrue}

\def\eqnarray{\stepcounter{equation}\let\@currentlabel\theequation
\global\@eqnswtrue\m@th
\global\@eqcnt\z@\tabskip\@centering\let\\\@eqncr
$$\halign to\displaywidth\bgroup\@eqnsel\hskip\@centering
     $\displaystyle\tabskip\z@{##}$&\global\@eqcnt\@ne
      \hfil$\;{##}\;$\hfil
     &\global\@eqcnt\tw@ $\displaystyle\tabskip\z@{##}$\hfil
   \tabskip\@centering&\llap{##}\tabskip\z@\cr}

\makeatother

\begin{document}
\thispagestyle{empty}
\vspace*{-15mm}
\baselineskip 10pt
\begin{flushright}
\begin{tabular}{l}
{\bf KEK-TH-555}\\
{\bf EPHOU-97-009}\\
{\bf hep-ph/9805447}\\
{\bf May 1998}
\end{tabular}
\end{flushright}
\baselineskip 24pt 
\vglue 15mm 
\begin{center}
{\Large\bf
Constraints on leptophobic $Z'$ models from 
electroweak experiments  
}
\vspace{5mm}

\baselineskip 18pt 
\def\thefootnote{\fnsymbol{footnote}}
\setcounter{footnote}{0}
{\bf
Yoshiaki Umeda$^{1,2}$\footnote{e-mail: umeda@theory.kek.jp},
Gi-Chol Cho$^{2}$\footnote{Research Fellow of the Japan Society 
for the Promotion of Science} and 
Kaoru Hagiwara$^{2}$

}
\vspace{5mm}

$^1${\it Department of Physics, Hokkaido University, Sapporo
060-0010, Japan} \\

$^2${\it Theory Group, KEK, Tsukuba, Ibaraki 305-0801, Japan}

\vspace{20mm}
\end{center}


\begin{center}
{\bf Abstract}\\[10mm]
\begin{minipage}{12cm}
\noindent

We study the constraints from updated electroweak data on the three 
leptophobic $Z'$ models, the ${\eta}$ model with an appropriate  
U$(1)'$-U$(1)^{}_Y$ kinetic mixing, a $Z'$ model motivated by the flipped 
SU(5) $\times$ U(1) unification, and the phenomenological $Z'$ model
of Agashe, Graesser, and Hinchliffe.
 The $Z$-$Z'$ mixing effects are parametrized in terms of a positive
contribution to the $T$ parameter, $T_{\rm new}$, and  
the effective mass mixing parameter, $\bar{\xi}$.
All the theoretical predictions for the $Z$ boson parameters,
the $W$ boson mass and the observables in low-energy 
neutral current experiments 
are presented together with the standard model radiative 
corrections. The allowed region in the ( $\bar{\xi},T_{{\rm new}}$) 
plane is shown for the three models.
The 95$\%$ CL lower limit on the heavier mass eigenstate 
$Z_2^{}$ is given as a function of the effective $Z$-$Z'$
mixing parameter $\zeta$.

\end{minipage}
\end{center}
\vspace{10mm}
\baselineskip 18pt 
{\small 
\begin{flushleft}
\end{flushleft}
}

\newpage
\baselineskip 16pt 
\def\thefootnote{\arabic{footnote}}
\setcounter{footnote}{0}
\section{Introduction}
Although the minimal standard model (SM) is in excellent 
agreement with current electroweak experiments~\cite{lep_slc_97}, 
it is still worth looking for 
evidences of physics beyond the SM at current or future 
experiments.
One of the simplest extensions of the SM is to introduce an 
extra U(1)~[$\equiv{\rm U(1)'}$] gauge boson in the  
weak scale. 
The presence of an additional U(1) gauge symmetry is expected, 
 e.g. in grand unified theories (GUTs) with a
gauge group whose rank 
is higher than that of the SM~(for a review of phenomenology, 
see \cite{hewett_rizzo,langacker}).
If an additional $Z'$ boson exists in the weak scale, then its 
properties are constrained 
from the electroweak experiments. 
Constraints on several $Z'$ models
have been obtained from 
recent data on the $Z$ pole experiments,
$W$ boson mass measurement, and the  
low-energy neutral current (LENC) experiments~\cite{langacker, chm}.

Among various models of additional $Z'$ bosons, leptophobic 
$Z'$ models~\cite{lepto_a, lepto_ET, lepto_Rb, eta_model, flipped, lbl} 
($Z'$ bosons whose couplings to leptons are 
either absent or negligibly small) are worth special 
attention because of the following reasons. First, such 
$Z'$ boson can have significant mixing with the SM $Z$ boson 
because the couplings of the observed $Z$ boson to quarks
are less precisely measured than those to leptons. Second, 
because most $Z'$ searches have so far been done at either 
purely leptonic channels 
($e^{+}_{}e^{-}_{} \rightarrow \ell^{+}_{}\ell^{-}_{}$) 
or lepton-quark processes 
($q\bar{q} \rightarrow \ell^{+}_{}\ell^{-}_{},  
e^{+}_{}e^{-}_{}\rightarrow q\bar{q}, 
 \ell q \rightarrow \ell q$, etc.)~\cite{L3ALEPH,CDFlq}, 
their lower mass bounds are much less stringent than those 
of the $Z'$ bosons with significant lepton couplings.
Because of the above properties, the consequences of leptophobic
$Z'$ bosons have often been studied when an experimental hint of 
non-standard model physics appeared: such as  
the excess of large $E_T$ jets at CDF~\cite{lepto_a,
lepto_ET, flipped} and 
the anomaly in the rates of $b$ or $c$ quark pair 
production at LEP1~\cite{lepto_a, lepto_ET, lepto_Rb, 
eta_model, flipped, lbl} .
Since the experimental data are now close 
to the SM predictions,  
phenomenological motivation of the leptophobic
$Z'$ models is weakened. 
But they still remain as one of the attractive new physics 
possibilities which may allow the existence of $Z'$ boson 
at accessible energy scale. 

In this paper, we study the constraints on some leptophobic $Z'$ 
models from the latest electroweak experiments. 
We will concentrate on the three models which have 
been proposed in association with the anomaly in the 
$Z \rightarrow b\bar{b},c\bar{c}$ widths, since the discrepancy 
in the $Z \rightarrow b\bar{b}$ sector have not been completely 
disappeared.
They are the models 
by Babu \etal~(model A~\cite{eta_model}), 
Lopez \etal~(model B~\cite{flipped}) 
and Agashe \etal~(model C~\cite{lbl}).
Leptophobia of the $Z'$ boson in these models is achieved;
(i) by an appropriate ${\rm U}(1)_Y \times {\rm U}(1)'$
kinetic mixing in the $\eta$-model of  
the supersymmetric E$_6$ model~\cite{hewett_rizzo} 
(model A~\cite{eta_model}) or 
(ii) by a suitable ${\rm U(1)'}$ charge assignment on 
the matter fields either in the flipped SU(5) $\times$
U(1) GUT framework (model B~\cite{flipped}), or in the 
effective low-energy gauge theory (model C~\cite{lbl}). 
In view of the latest electroweak experiments, we would
like to re-examine the impacts of these models quantitatively
and update the mass bounds of the $Z'$ bosons. 

These three models should have several extra matter fields 
in the electroweak scale in order 
to cancel the ${\rm U(1)'}$ gauge anomaly. 
Furthermore, the models A and B are supersymmetric and they 
contain supersymmetric particles in their spectrum. 
We assume that all the extra particles are heavy enough to
decouple from the SM radiative corrections and that 
they do not give rise to new low-energy interactions
among quarks and leptons.

This paper is organized as follows. 
In the next section, we briefly review the characteristic 
feature of general \zzp mixing in order to fix our notation. 
We show that the extra contributions to the neutral current 
interactions due to the \zzp mixing can be parametrized 
in terms of a positive contribution to the 
$T$-parameter~\cite{peskin_takeuchi}, $T_{\rm new}$, 
and the effective $Z$-$Z'$ mass mixing angle, $\ov{\xi}$. 
In Section 3, we review the updated LEP or SLAC Linear
Collider SLC data reported 
at the summer conferences in 1997~\cite{lep_slc_97} and 
the low-energy data~\cite{chm} which are used in our fit. 
We also present the theoretical predictions of the electroweak
observables in the three 
leptophobic $Z'$ models by using the parameters 
$T_{\rm new}$ and  $\ov{\xi}$,
together with the SM radiative corrections~\cite{hhkm}. 
Constraints on the three leptophobic models under 
the current experimental data are then given in Section 4. 
The lower mass limits of the $Z'$ boson for three models are also given.
Section 5 is devoted to summarize our results. 
\section{Brief review of generalized \zzp mixing}
\clean
In this section, we give a brief review of the phenomenological 
consequences of general $Z'$ models. 
Following the focus of this paper, we concentrate on the \zzp 
mixing and the neutral current interactions. 

If an additional gauge boson ($Z'$) couples to the SM  
Higgs field whose vacuum expectation value (VEV) induces 
the spontaneous symmetry breaking of the SM gauge symmetry, 
SU$(2)_L \times$ U$(1)_Y$, the \zzp 
mass mixing term appears after 
the Higgs field develops the VEV. 
In addition, the kinetic mixing between the $Z'$ and the 
hypercharge gauge boson $B$ can occur at high energy scale~\cite{holdom}.
The effective Lagrangian of the neutral gauge bosons then takes 
the following general form~\cite{holdom}:

\bea
{\cal L}_{gauge} 
	&=&  -\frac{1}{4}Z^{\mu\nu}Z_{\mu\nu}
            -\frac{1}{4}Z'^{\mu\nu}Z'_{\mu\nu} 
	    -\frac{\sin \chi}{2}B^{\mu\nu}Z'_{\mu\nu}
	    -\frac{1}{4}A^{0\mu\nu}A^{0}_{\mu\nu} 
\nonumber \\ 
	& & + m^2_{ZZ'} Z^{\mu}Z'_{\mu}
	    +\frac{1}{2} m^2_Z Z^{\mu}Z_{\mu}
	    +\frac{1}{2} m^2_{Z'} Z'^\mu Z'_{\mu}.
\label{eq:l_gauge}
\eea
Here the term proportional to  $\sin \chi$
denotes the kinetic mixing and the term proportional to
$m^2_{ZZ'}$ denotes the mass mixing.
The kinetic and mass mixings can be removed by the following 
transformation from the current eigenstates $(Z,Z',A^0)$ 
to the mass eigenstates $(Z_1,Z_2,A)$:
\begin{equation}
\left( \begin{array}{c}  Z \\  Z' \\  A^0  
\end{array}\right) 
= 
\left(
	\begin{array}{ccc}
\cos \xi + \sin \xi \sin \theta_W \tan \chi &
-\sin \xi + \cos \xi \sin\theta_W \tan \chi & 0 \\
\sin \xi / \cos \chi & \cos \xi / \cos \chi & 0 \\
-\sin\xi \cos \theta_W \tan \chi & 
- \cos \xi \cos \theta_W \tan \chi & 1   
	\end{array}
\right) 
\left( \begin{array}{c} {Z_1} \\ 
{Z_2} \\ {A} \end{array}\right).  
\end{equation}
Here the mixing angle $\xi$ is given by 
\begin{equation}
\tan 2\xi = \frac{-2c^{}_{\chi}(m^2_{ZZ'}+s^{}_W s^{}_{\chi}
            m^2_Z)}{m^2_{Z'} - (c^2_{\chi}-s^2_W s^2_{\chi})m^2_Z+
            2s^{}_W s^{}_{\chi} m^2_{ZZ'}}, 
\label{eq:angle_xi}
\end{equation}
with the short handed notation, 
$c_\chi = \cos\chi$, $s_\chi = \sin\chi$ 
and $s_W = \sin\theta_W$. 
The physical masses $m_{Z_1}$ and $m_{Z_2}$ ($m_{Z_1} < m_{Z_2}$)  
are given as follows: 
\bsub
\bea
m_{Z_{1}}^2 &=& 
  m_Z^2 (c_\xi + s_\xi s_W t_\chi)^2 
+ m_{Z'}^2 \biggl( \frac{s_\xi}{c_\chi} \biggr)^2
+ 2 m^2_{ZZ'} \frac{s_\xi}{c_\chi} (c_\xi + s_\xi s_W t_\chi), 
\\
m_{Z_{2}}^2 &=& 
  m_Z^2 (c_\xi s_W t_\chi - s_\xi)^2 
+ m_{Z'}^2 \biggl( \frac{c_\xi}{c_\chi} \biggr)^2
+ 2 m^2_{ZZ'} \frac{c_\xi}{c_\chi} (c_\xi s_W t_\chi -s_\xi), 
\eea
\esub
where $c_\xi = \cos\xi$, $s_\xi = \sin\xi$ and 
$t^{}_{\chi}$ = $\tan\chi$.

Except in the limit of the perfect leptophobity and the 
weak coupling,
the observed $Z$ boson should be identified with the 
lighter mass eigenstate, the $Z_1$ boson.
Generally, the good agreement between the current 
experimental results and the SM predictions 
then implies that the angle $\xi$ have to be small.
In the small $\xi$ limit, the interaction Lagrangian of
the $Z_1$ boson couplings to quarks can be expressed as 
\begin{eqnarray}
{\cal L}_{Z_1} &\!=\!& -g_Z Z_{1}^{ \mu}\!
	\left[ \overline{u_L}\gamma_{\mu}
	\!\left( \frac{1}{2}-\frac{2}{3} \sin^2\theta_W  
	+ Q^{Q}_E \xibar \right) u_L + \overline{d_L}\gamma_{\mu} 
	\!\left(-\frac{1}{2}+\frac{1}{3}\sin^2\theta_W 
	+ Q^{Q}_E \xibar \right)\! d_L \right. 
\nonumber \\ 
	&&\ \left. +\overline{u_R} \gamma_{\mu} 
	\!\left(-\frac{2}{3}\sin^2\theta_W + Q^U_E \xibar \right)\!u_R 
	+ \overline{d_R}\gamma_{\mu}
	\!\left(\frac{1}{3}\sin^2\theta_W + Q^D_E \xibar \right)\!
	 d_R\right]. 
\label{eq:neutral}
\end{eqnarray}
In the above, the effective ${\rm U(1)'}$ charge $Q_E^f (f=Q,U,D)$ 
is given as 
\bsub
\bea
Q^f_E &=& Q'^f_E + Y_f \delta, 
\label{eq:effective_charge}\\
\delta &=&  -\frac{g^{}_Z}{g^{}_E}s^{}_W s^{}_{\chi}, 
\eea
\label{eq:u1_charge}
\esub
$\!\!$where $Y_f$ in eq.~(\ref{eq:effective_charge}) 
represents the hypercharge of the quark multiplet $f$ and $Q'^f_E$
represents the $Z'$ coupling;
\begin{equation}
{\cal L}_{Z'} = -g_E^{} Z'_{\mu} \sum_f Q^{\prime f}_E
(\overline{f} \gamma^{\mu} f).
\end{equation}
Here $Q=(u^{}_L, d^{}_L)^{\rm T}$ denotes the left-handed 
quark doublet and $U = u^{}_R$,
$D = d^{}_R$ denote the right-handed quark singlets.
The effective mixing parameter $\xibar$ 
is related to the mixing angle $\xi$~(\ref{eq:angle_xi}) by 
\beq
\xibar = \frac{g_E}{g_Z\cos \chi }\xi.
\eeq

The requirement that the extra contributions to the 
leptonic neutral current are zero is expressed simply as
\beq
Q_E^\ell = Q_E'^\ell + Y_\ell \delta = 0\;\;\;\; 
{\rm for} \; \ell = L \; {\rm and} \; E,
\label{eq:lepty}
\eeq
where $L=(\nu^{}_L,\ell^{}_L)^{\rm T}\; (\ell = e,\mu,\tau)$
is the left-handed lepton doublet and $E = \ell^{}_R$ is
the right-handed lepton singlet.
In the model A~\cite{eta_model}, the $Z'$ coupling $Q^{\prime \ell}_E$
is proportional to $Y_{\ell}$ and the 
leptophobia is achieved by taking $\delta 
= 1/3$. In the model B~\cite{flipped}, the $Z'$ boson couples only
to the decouplet of flipped SU(5)$\times$U(1) group, which consist of
the quark doublet, $Q$, the down-quark singlet $D^c$ and 
the right-handed neutrino. The leptophobity condition~(\ref{eq:lepty})
is hence satisfied when the kinetic mixing is suppressed 
($\delta = 0$) by a certain discrete symmetry~\cite{flipped}.
Finally in the model C~\cite{lbl}, the ratios of the tree quark
couplings $Q^Q_E$, $Q^U_E$, $Q^D_E$, are fixed by 
phenomenologically by setting $Q^L_E$ = $Q^E_E$ = 0. We set 
$\delta = 0$ in the following analysis of the model C.
The ${\rm U(1)'}$ charge assignments of quarks 
in the three models are summarized in Table~\ref{tab:charge}. 
\extra_charge 

The presence of the mass shift affects the 
electroweak $T$-parameter~\cite{peskin_takeuchi}. Following 
the notation of ref.~\cite{hhkm}, the $T$-parameter 
is expressed in terms of the effective form factors $\gzbar(0), 
\gwbar(0)$ and the fine structure constant $\alpha$:
\bsub
\bea
\alpha T &\equiv& 1 - \frac{\bar{g}^2_{W}(0)}{m^2_W}
	\frac{m^2_{Z_1}}{\bar{g}^2_Z(0)}  
\label{eq:Tparaa}\\ 
	&=&
	\alpha \left(T_{\rm SM}^{}+  T_{\rm new}^{}\right), 
\eea
\label{eq:Tpara}
\esub
$\!\!$where $m^{}_{Z_1}$ is the $Z$-boson mass measured precisely 
at LEP1. The SM contribution $T_{\rm SM}^{}$ and the 
new physics contribution $T_{\rm new}$ are each expressed as: 
\bsub
\bea
\alpha T_{\rm SM} 
	& = & 1 - \frac{\bar{g}^2_{W}(0)}{m^2_W}
		\frac{m^2_{Z}}{\bar{g}^2_Z(0)}, 
\label{eq:TSMa} \\
\alpha T_{\rm new}
	& = & -\frac{ \Delta m^2}{m^2_{Z_1}} \geq 0.
\label{eq:TSMb}
\eea
\label{eq:TSM}
\esub
$\!\!$It should be noted that the parameter $T_{\rm SM}$~(\ref{eq:TSMa}) 
is calculable in the SM as a function of $m^{}_t$ and 
$m^{}_H$, whereas the precision electroweak experiments 
measure the $T$-parameter~(\ref{eq:Tparaa}).
Hence by using the observed $m^{}_t$~\cite{cdfd0}
and the theoretical/experimental constraints on $m^{}_H$, we can 
obtain the constraint on $T_{\rm new}$.
It should also be noticed here that the contribution of 
the mass shift term to the $T$-parameter, $T_{\rm new}$, 
is always positive whether the kinetic mixing is 
presented or not. 
We note here that the effect of the kinetic mixing term 
proportional to the hypercharge $Y_f$ can be absorbed by 
further re-defining the $S$ and $T$ 
parameters~\cite{holdom2,general_zzmixing}.
However, we find no merit in doing so when the $Z'$ charges
$Q^{\prime f}_E$ are not negligible as compared to the 
$Y_f \delta$ terms in eq.~(\ref{eq:effective_charge}).
We therefore retain the $Y_f \delta$ terms as parts of 
the explicit contribution to the $Z_1$ couplings.
Physical consequences of the models are of course independent 
of our choice of parametrization.
The consequences of the $Z$-$Z'$ mixing can therefore be 
parametrized solely by the charges $Q^{f}_E$ and the 
two parameters $\bar{\xi}$ and $T_{\rm new}$.
In the following analysis, we find that the parameter 
\begin{equation}
\zeta = \frac{g^{}_Z}{g^{}_E}\frac{m^2_{ZZ'}}{m^2_Z}-\delta,
\label{eq:zeta}
\end{equation}
plays an essential role in determining the $Z$-$Z'$ mixing 
phenomenology. It can be calculated in a given gauge model 
whose gauge couplings are unified at a certain high energy scale, 
once the Higgs sector and the matter particles are fixed.
At $\zeta=0$, the $Z$-$Z'$ mixing disappears; see 
eq.~(\ref{eq:angle_xi}). In the small mixing limit, we find the
following parametrizations useful:
\begin{subequations}
\begin{eqnarray}
\bar{\xi} &=& -\left(\frac{g^{}_E}{g^{}_Z}
\frac{m^{}_Z}{m^{}_{Z'}}\right)^2\zeta
\left[1+O(\frac{m^2_Z}{m^2_{Z'}})^2\right], \\
\alpha T_{\rm new} &=& -\frac{\Delta m^2}{m^2_{Z'}} = 
\left(\frac{g_E^{}}{g_Z^{}}\frac{m^{}_Z}{m^{}_{Z'}}\right)^2\zeta^2
\left[1+O(\frac{m^2_Z}{m^2_{Z'}})^2\right].
\end{eqnarray}
\label{eq:approx_Txi}
\end{subequations}
\lep_table
\section{Electroweak observables in the leptophobic $Z'$ models}
\clean

In this section, we present the theoretical prediction of the 
electroweak observables which are used in our analysis.  
The results are parametrized in terms of the SM parameters,
$m_t$, $m^{}_H$, $\alpha_s(m^{}_{Z_1})$ and 
$\bar{\alpha} (m^2_{Z_1})$, and 
the two new physics parameters, $\bar{\xi}$ and $T_{\rm new}$.
We collect in Table~\ref{tab:lep_table} the updated results of 
the electroweak measurement, reported at the summer conferences 
in 1997~\cite{lep_slc_97}. 
Besides the $Z$-pole experiments, we also include in our fit 
the results of some low-energy neutral current 
experiments~\cite{chm}, which are 
also listed in Table~\ref{tab:lep_table}. 
\subsection{$Z$ boson parameters}
The decay amplitude for the process 
$Z^{}_1 \to f^{}_{\alpha}\overline{f^{}_{\alpha}}$ 
is written as
\begin{equation}
T(Z_1 \rightarrow f_{\alpha} \overline{f_{\alpha}}) = 
M^f_{\alpha} \;\epsilon_{Z_1} \cdot J_{f_{\alpha}}, 
\label{eq:decay_amp}
\end{equation}
where $f^{}_{\alpha}$ stand for the leptons and quarks, and  
$\alpha=L,R$ denotes their chirality. 
$\epsilon_{Z_1}^\mu$ is the polarization vector of the
$Z^{}_1$ boson and 
$J^{\mu}_{f_{\alpha}}=\overline{f_{\alpha}}\gamma^{\mu}f_{\alpha}$ 
is the fermion current without the coupling factors. 
Since all the pseudo-observables of the  $Z$-pole experiments can be expressed 
in terms of the real scalar amplitude $M_\alpha^f$ of 
eq.~(\ref{eq:decay_amp}), it is useful to 
present their theoretical predictions. By adopting the 
notation~\cite{lep_slc_97}
\begin{equation}
g^f_{\alpha} = \frac{M^f_{\alpha}}{\sqrt{4\sqrt{2}G_Fm^2_{Z_1}}} 
\approx \frac{M^f_{\alpha}}{0.74070},
\end{equation}
all the theoretical predictions can be expressed in the form 
\begin{equation}
g^f_{\alpha} = (g^f_{\alpha})^{}_{\rm SM}+Q^{f_{\alpha}}_E \bar{\xi}.
\end{equation}
When the $Z'$ couplings are independent of the fermion 
generation,\footnote{In the string-inspired flipped SU(5)$\times$U(1)
models~\cite{flipped}, the generation-independence of the $Z'$
couplings does not necessarily hold. We study in this paper 
only the generation independent case for brevity.}
the following eight amplitudes determine all the $Z$-pole (-pseudo)
observables:
\begin{subequations}
\begin{eqnarray}
g^{\nu}_L &=& \makebox[3.3mm]{ } 0.50214 
	+ 0.453\,{\Delta \bar{g}^2_Z},
\label{eq:amp_nu} \\
g^e_L &=& - 0.26941 - 0.244\,{\Delta \bar{g}^2_Z} 
	+ 1.001\,{\Delta \bar{s}^2},  
\label{eq:amp_el}\\
g^e_R &=& \makebox[3.3mm]{ } 0.23201 + 0.208\,
	{\Delta \bar{g}^2_Z} + 1.001\,{\Delta \bar{s}^2}, 
\label{eq:amp_er}\\
g^u_L &=&  \makebox[3.3mm]{ } 0.34694 + 0.314\,
	{\Delta \bar{g}^2_Z} - 0.668\,{\Delta \bar{s}^2}
	+ \left( \begin{array}{r} - 0.278 \\  0.278 \\ - 0.278 
	\end{array} \right) \,{\it \bar{\xi}}, 
\label{eq:amp_ul}\\
g^u_R &=& - 0.15466 - 0.139\,{ \Delta \bar{g}^2_Z}  
	- 0.668 \,{\Delta \bar{s}^2} + \left( \begin{array}{r} 
	0.556 \\ 0 \\ \makebox[3.3mm]{ } 0.556 
	\end{array} \right) \,{\it \bar{\xi}},  
\label{eq:amp_ur}\\
g^d_L &=& - 0.42451 - 0.383\,{\Delta \bar{g}^2_Z} 
	+ 0.334\,{\Delta \bar{s}^2} +\left( \begin{array}{r}
	- 0.278 \\ 0.278 \\ -0.278
	\end{array} \right) \,{\it \bar{\xi}}, 
\label{eq:amp_dl}\\
g^d_R &=& \makebox[3.3mm]{ }  0.07732
	+ 0.069\,{ \Delta \bar{g}^2_Z} + 0.334\,{\Delta \bar{s}^2}  
	+ \left( \begin{array}{r}
	- 0.278 \\ - 0.278 \\ 0.278
	\end{array} \right) \,{\it \bar{\xi}},  
\label{eq:amp_dr}\\
g^b_L &=& - 0.42109 - 0.383\,{ \Delta \bar{g}^2_Z} 
	+ 0.334\,{\Delta \bar{s}^2} + \left( \begin{array}{r}
	- 0.278 \\ 0.278 \\ - 0.278
	\end{array} \right) \,{\it \bar{\xi}} \nonumber  \\
	& & + 0.00043 x_t, 
\label{eq:amp_bl}
\end{eqnarray}
\label{eq:amp}
\end{subequations}
$\!\!$where the absence of terms proportional to $\xibar$ in 
eqs.~(\ref{eq:amp_nu})$\sim$(\ref{eq:amp_er}) is the consequence of 
the leptophobia of the $Z'$ component of the mass eigenstate $Z_1$. 
The first, the second and the third rows of the 
column vector multiplying $\bar{\xi}$ 
in eqs.~(\ref{eq:amp_ul})$\sim$(\ref{eq:amp_bl}) give the $Q^f_E$ 
charges of the models A, B and C, respectively. 
The terms $\Delta \bar{g}^2_Z$ and $\Delta \bar{s}^2$ denote
the shift in the effective coupling, $\gzbar(m^2_{Z_1})$ and 
$\sbar(m^2_{Z_1})$~\cite{hhkm} from their reference values at 
$m^{}_t$ = 175 GeV, $m^{}_H$ = 100 GeV, 
$1/\bar{\alpha}(m^2_{Z_1})$ = 128.75:
\bsub
\begin{eqnarray}
\hspace{-10mm}
\Delta \bar{g}^2_Z &=& \bar{g}^2_Z(m^2_{Z_1})-0.55635 =  0.00412 \Delta T + 
	0.00005[1-(100{\rm \;GeV}/m^{}_H)^2], \\  
\hspace{-10mm}
\Delta \bar{s}^2 &=& \bar{s}^2(m^2_{Z_1})-0.23035 = 0.00360 \Delta S
			- 0.00241 \Delta T - 0.00023 \xa. 
\end{eqnarray}
\esub
Here $\Delta S$, $\Delta T$ and $\Delta U$ are also measured 
from their reference SM values and they can be expressed as 
the sum of the SM contribution and the new physics contribution: 
\begin{equation}
\Delta S = \Delta S^{}_{\rm SM}+S^{}_{\rm new}, \;\;
\Delta T = \Delta T^{}_{\rm SM}+T^{}_{\rm new}, \;\; 
\Delta U = \Delta U^{}_{\rm SM}+U^{}_{\rm new}.
\label{eq:delSTU}
\end{equation}
The SM contributions are parametrized as~\cite{hhm} 
\begin{subequations}
\begin{eqnarray}
\Delta S^{}_{\rm SM} &=& - 0.007 x^{}_t +0.091 x^{}_H -0.010 x^2_H , \\
\Delta T^{}_{\rm SM} &=& (0.130 - 0.003 x^{}_H) x^{}_t +0.003 x^{2}_t 
- 0.079 x^{}_H -0.028 x^2_H \nonumber \\ & & +0.0026 x^3_H, \\
\Delta U^{}_{\rm SM} &=& 0.022 x^{}_t -0.002x^{}_H,
\end{eqnarray}
\end{subequations}
in terms of the parameters 
\begin{subequations}
\begin{eqnarray}
\xh &\equiv& \log (\mh/100{\rm \;GeV}),
\label{eq:xh} \\
\xa &\equiv& \frac{1/\abar(m^2_{Z_1})-128.75}{0.09}, \\ 
\xt &\equiv& \frac{\mt - 175{\rm \;GeV}}{10{\rm \;GeV}}, 
\end{eqnarray}
\end{subequations}
which measure the deviations of the SM parameters from their reference
values. The $m^{}_t$-dependence of the $Zb^{}_Lb^{}_L$
vertex correction is given explicitly by the last term 
proportional to $x^{}_t$ in eq.~(\ref{eq:amp_bl}).

In term of the above eight effective amplitudes, all the partial width
of the $Z_{1}$ boson can be expressed as 
\begin{equation}
\Gamma_f = \frac{G_Fm_{Z_1}^{3}}{3\sqrt{2} \pi} \left\{ 
	\left| g^f_V \right|^2\frac{C_{fV}}{2} 
	+ \left| g^f_A \right|^2 \frac{C_{fA}}{2}  \right\}
	\left( 1+\frac{3}{4}Q^2_f\frac{\bar{\alpha}(m^2_{Z_1})}{\pi}\right)
        \makebox[10mm][l]{,}
\label{eq:partial_width}
\end{equation}
where 
\begin{equation}
g^f_V = g^f_L + g^f_R,\;\;\;
g^f_A = g^f_L - g^f_R.
\end{equation}
The factors $C_{fV}, C_{fA}$ account for the
finite mass corrections and the final state QCD 
corrections for quarks.
Their numerical values are listed in Table~\ref{tab:cvca}.
The $\alpha_s$-dependence of the QCD corrections is given in terms of
the parameter
\begin{equation}
x_s^{} \equiv \frac{\alpha_s(m_{Z_1}) - 0.118}{0.003}. 
\end{equation}
The last term proportional to $\abar(m^2_{Z_1})/\pi$ 
in the partial decay width $\Gamma_f$~(\ref{eq:partial_width}) 
accounts for the final state QED correction. 
\cvca_tab

In terms of the partial widths, the hadronic width $\Gamma_h$
and the total width $\Gamma_{Z_1}$ are obtained as
\bsub
\begin{eqnarray}
\Gamma_h &=& \Gamma_u + \Gamma_c + \Gamma_d + \Gamma_s + \Gamma_b, 
\label{eq:hadron_width} \\
\Gamma_{Z_1} &=& 3\Gamma_{\nu} + \Gamma_e 
	+ \Gamma_{\mu} + \Gamma_{\tau} + \Gamma_h.
\label{eq:total_width}
\end{eqnarray}
\esub
The ratios $R_\ell^{}, R_c^{}, R_b^{}$ of the branching fractions
and the hadronic peak cross section $\sigma_h^0$ can now
be obtained from
\begin{equation}
R_{\ell} = \frac{\Gamma_h}{\Gamma_{\ell}},\;
R_c      = \frac{\Gamma_c}{\Gamma_h},\;
R_b      = \frac{\Gamma_b}{\Gamma_h},\;
\sigma^0_h = \frac{12\pi}{m^2_{Z_1}}
	\frac{\Gamma_e\Gamma_h}{\Gamma_{Z_1}^2}. 
\end{equation}

The left-right asymmetry parameters $A^f$ are also obtained from
the effective amplitudes~(\ref{eq:amp}) by
\begin{equation}
A^f = \frac{(g^{f}_L)^2-(g^{f}_R)^2}{(g^{f}_L)^2+(g^{f}_R)^2} = 
\frac{2g^{f}_V g^{f}_A}{(g^{f}_V)^2+(g^{f}_A)^2}, 
\end{equation}
in terms of which all the $Z$-pole asymmetry data are constructed.
The forward-backward~(FB) asymmetry $A^{0,f}_{FB}$ 
and the left-right~(LR) asymmetry $A^{0,f}_{LR}$ are
\bsub
\bea
A^{0,f}_{FB} &=& \frac{3}{4}A^{e}A^{f}, \\
A^{0,f}_{LR} &=&  A^{f}.
\eea
\esub
It is now straightforward to obtain the predictions for 
all the $Z$-pole observables in the presence of the $Z$-$Z'$
mixing. We list below the compact parametrizations for all
the $Z$-pole observables of Table~\ref{tab:lep_table}.
The leptonic and hadronic decay widths are 
\begin{subequations}
\begin{eqnarray}
\Gamma_{\nu} &=& 0.16730  + 0.302 {\Delta \bar{g}^2_Z }, \\
\Gamma_{e}   &=& 0.08403  + 0.152\,{\Delta \bar{g}^2_Z } 
		- 0.050  \,{\Delta \bar{s}^2}, \\
\Gamma_{h}   &=& 1.7434 + 3.15 {\Delta \bar{g}^2_Z} 
		- 2.50\,{\Delta \bar{s}^2} 
		+ 0.0017\,x^{\prime}_s  
		+ \left( \begin{array}{r}
		-0.33 \\ -0.91 \\ 0.19
		\end{array} \right) \,{\it \bar{\xi}}, 
\end{eqnarray}
\end{subequations}
where the parameter $x_s^{\prime}$ 
\begin{equation}
x_s^{\prime} = x_s^{}-0.44 x_t^{},
\end{equation}
gives the combination of $\alpha_s(m^{}_{Z_1})$ and the $Zb^{}_Lb^{}_L$
vertex correction~\cite{hhm} which is constrained by the 
$Z$ boson hadronic width.
The total decay width, the ratios of the partial 
widths and the hadronic peak cross section are 
\bsub
\bea
\Gamma_Z &=& 2.9977{\Gamma_{e}}+3{\Gamma_{\nu}}+{\Gamma_{h}}
\nonumber \\
	&=& 2.4972 + 4.51  \,{\Delta \bar{g}^2_Z} 
	- 2.65\,{\Delta \bar{s}^2} 
	+ 0.0017\,x_s^{\prime} 
	+ \left( \begin{array}{r}
	-0.33 \\ -0.91 \\ 0.19
	\end{array} \right) \,{\bar{\xi}}, \\ 
R_l &=& 20.747 + 0.05\,{\Delta \bar{g}^2_Z} - 17.42 \,{\Delta \bar{s}^2} 
	+ 0.020\,x_s^{\prime}
	+ \left( \begin{array}{r}	 -3.9 \\  -10.8 \\ 2.3
	\end{array} \right) \,{\it \bar{\xi}}, \\
R_c &=& 0.1721 - 0.0004\,{\Delta \bar{g}^2_Z} - 0.058\,{\Delta \bar{s}^2} 
	+ \left( \begin{array}{r}  -0.40 \\  0.32 \\ -0.45
	\end{array} \right) \,{\it \bar{\xi}},  \\
R_b &=& 0.2157 + 0.002 \,{\Delta \bar{g}^2_Z} + 0.036 \,{\Delta \bar{s}^2}
	- 0.0003\,x_t 
	+ \left( \begin{array}{r}	0.27 \\ -0.21 \\ 0.30
	\end{array} \right) \,{\it \bar{\xi}}, \\ 
\sigma_h^0 &=& 41.474 + 0.01\,{\Delta \bar{g}^2_Z} 
	+ 3.92\,{\Delta \bar{s}^2} - 0.016\,x_s^{\prime} 
	+ \left( \begin{array}{r} 3.1 \\ 8.6 \\ -1.8
	\end{array} \right) \,{\it \bar{\xi}}. 
\eea
\label{eq:grs}
\esub
The asymmetry parameters are  
\bsub
\bea
A^{0,l}_{FB} &=& 0.0165 + 0.002\,{\Delta \bar{g}^2_Z} 
	- 1.75 \,{\Delta \bar{s}^2}, \\
A^{0,c}_{FB} &=& 0.0744 + 0.004\,{\Delta \bar{g}^2_Z} 
	- 4.32\,{\Delta \bar{s}^2} 
	+ \left( \begin{array}{r}
	0.17 \\ 0.05 \\ 0.17
	\end{array} \right) \,{\it \bar{\xi}}, \\
A^{0,b}_{FB} &=& 0.1040 + 0.005\,{\Delta \bar{g}^2_Z} 
	- 5.58\,{\Delta \bar{s}^2 } 
	+ \left( \begin{array}{r}
	0.06 \\ 0.04 \\ -0.04
	\end{array} \right) \,{\it \bar{\xi}},  \\
A^{e}_{LR} &=& 0.1484 + 0.007 \,{\Delta \bar{g}^2_Z} 
	- 7.86\,{\Delta \bar{s}^2}, \\
A_{c} &=& 0.668 + 0.003\,{\Delta \bar{g}^2_Z} - 3.45\,{\Delta \bar{s}^2}  
	+ \left( \begin{array}{r}
	1.5 \\ 0.4 \\ 1.5
	\end{array} \right) \,{\it \bar{\xi}}, \\
A_{b} &=& 0.935 + 0.001\,{\Delta \bar{g}^2_Z} - 0.65\,{\Delta \bar{s}^2} 
	+ \left( \begin{array}{r}
	0.54 \\ 0.37 \\ -0.37
	\end{array} \right) \,{\it \bar{\xi}}. 
\end{eqnarray}
\label{eq:asmfl}
\end{subequations}

Eqs.~(\ref{eq:grs}) and ~({\ref{eq:asmfl}) give all the $Z$-pole 
observables in terms of the four SM parameters
$m_t$, $m^{}_H$, $\alpha_s(m^{}_{Z_1})$ and $\bar{\alpha}(m^2_{Z_1})$, 
and the two new physics parameters $\bar{\xi}$ and $T_{\rm new}$,
by setting $S_{\rm new}$ = $U_{\rm new}$ = 0 in eq.~(\ref{eq:delSTU}).
The latter condition and the expression ~(\ref{eq:TSMb}) hold when 
all the exotic particles obtain the large 
SU$(2)_L$ $\times$ U$(1)_Y$ invariant masses.
\subsection{$W$ boson mass}
The theoretical prediction of $m^{}_W$
can be parametrized by $\Delta S$, $\Delta T$, $\Delta U$
and $x^{}_{\alpha}$~\cite{hhm}: 
\begin{equation}
m_W^{}{\rm (\;GeV)} =80.402-0.288\,{\it \Delta S}+0.418\,{\it \Delta T} 
+0.337\,{\it \Delta U}+0.012\,{\it x_{\alpha}}. 
\end{equation}
\subsection{Observables in low-energy experiments}
In this subsection, we show the theoretical predictions 
for the electroweak observables in the low-energy neutral current 
(LENC) experiments --- 
(i) polarization asymmetry of the charged lepton scattering off 
nucleus target (3.3.1-- 3.3.4), 
(ii) parity violation in cesium atom (3.3.5) 
and (iii) inelastic $\nu_\mu$ scattering off nucleus target 
(3.3.6). 
They can be parametrized by $\Delta S$ and $\Delta T$~\cite{chm}.
The experimental results of (i), (ii) are given in terms of 
the model-independent parameters $C_{1q}, C_{2q}$~\cite{jekim} 
and $C_{3q}$~\cite{chm}. 
Those of (iii) are given by the parameters 
~$q_\alpha (q=u,d$ and $\alpha = L,R)$. 
All these model-independent parameters are 
expressed in terms of 
the helicity amplitudes of the low-energy lepton-quark scattering 
and their explicit forms are found in ref.~\cite{hhkm}.
In the following, we neglect the $Z_2$-contributions which are
doubly suppressed by the small mixing angle $\xi$ and the ratio
$m^2_{Z_1}/m^2_{Z_2}$ in the leptophobic $Z'$-models.
\subsubsection{SLAC $e$D experiment}
The parity asymmetry in the inelastic scattering of polarized 
electrons from the deuterium target has been measured at 
SLAC~\cite{slac}. 
The experiment was performed at the mean momentum transfer
$\langle Q^2 \rangle = $ 1.5 GeV$^2$. 
The experiment constrains the parameters  
$2C_{1u}-C_{1d}$ and $2C_{2u}-C_{2d}$. 
The most stringent constraint is found for the following combination
\begin{subequations}
\begin{eqnarray}
A_{\rm SLAC} &=& 2C_{1u}-C_{1d} +0.206(2C_{2u}-C_{2d}) \\
	&=& 0.745 - 0.016\,{\it \Delta S} + 0.016\,{\it \Delta T}
	+ \left( \begin{array}{r}
	 1.10 \\ 0.56 \\  0.55
	\end{array} \right) \,{\it \bar{\xi}}.  
\end{eqnarray}
\label{eq:slac_sm1}
\end{subequations}
\subsubsection{CERN $\mu^\pm$C experiment}
The CERN $\mu^\pm$C experiment~\cite{cern} has measured 
the charge and polarization asymmetry of deep-inelastic 
muon scattering off the ${}^{12}$C target.
The mean momentum transfer of the experiment may be estimated 
as $\langle Q^2 \rangle = $ 50 GeV$^2$~\cite{souder}.
The experiment constrains 
the parameters $2C_{2u}-C_{2d}$ and 
$2C_{3u}-C_{3d}$. 
The most stringent constraint is found for the following combination
\begin{subequations}
\begin{eqnarray}
A_{\rm CERN} &=& 2C_{3u}-C_{3d}+0.777(2C_{2u}-C_{2d}) \\
	 &=& -1.42-0.016\,{\it \Delta S}+0.0006\,{\it \Delta T}  
	+ \left( \begin{array}{r}
	1.58 \\  0 \\ 1.06
	\end{array} \right) \,{\it \bar{\xi}}.  
\end{eqnarray}
\label{eq:cern_sm1}
\end{subequations}
\subsubsection{Bates $e$C experiment} 
The polarization asymmetry of the electron elastic scattering
off the ${}^{12}$C target has been measured at Bates \cite{bates}.
The experiment was performed at the mean momentum transfer
$\langle Q^2 \rangle = $ 0.0225 GeV$^2$.
The experiment constrains a combination $C_{1u}+C_{1d}$. 
The theoretical prediction is
\begin{subequations}
\begin{eqnarray}
A_{\rm Bates} &=& C_{1u}+C_{1d} \\
	&=& - 0.1520 - 0.0023\,{\it \Delta S} 
	+ 0.0004\,{\it \Delta T} + \left( \begin{array}{r}
	-0.28 \\ 0.28 \\ 0.28
	\end{array} \right) \,{\it \bar{\xi}}. 
\end{eqnarray}
\end{subequations}
\subsubsection{Mainz $e$Be experiment }
The polarization asymmetry of electron quasi-elastic scattering
off the ${}^9$Be target has been measured at Mainz \cite{mainz}.
The experiment was performed at the mean momentum transfer
$\langle Q^2 \rangle = $ 0.2025 GeV$^2$. 
The asymmetry parameter $A_{\rm Mainz}$ is measured and its
theoretical prediction is
\begin{subequations}
\begin{eqnarray}
A_{\rm Mainz} &=& -2.73 C_{1u} + 0.65 C_{1d} - 2.19 C_{2u} 
	+ 2.03 C_{2d} \\ 
        &=&  -0.876 + 0.043\,{\it \Delta S} 
	-0.035\,{\it \Delta T}
	+ \left( \begin{array}{r}
	- 1.03 \\ - 0.74  \\ -0.73
	\end{array} \right) \,{\it \bar{\xi}}. 
\end{eqnarray}
\end{subequations}
\subsubsection{Atomic Parity Violation}
The experimental results of parity violation in atomic physics 
are often given in terms of the weak charge $Q^{}_W(A,Z)$
of nuclei. Using the model-independent parameters $C_{1q}^{}$,
the weak charge can be expressed as
\begin{equation}
Q^{}_W(A,Z)=2ZC_{1p}^{}+2(A-Z)C_{1n}^{}, 
\end{equation}
where the parameters $C_{1p}$ and $C_{1n}$ are obtained
from the quark-level amplitudes $C_{1u}$ and $C_{1d}$
after taking into account the hadronic corrections~\cite{marciano}.
The theoretical predictions for $C_{1p}$ and $C_{1n}$ are
\begin{subequations}
\begin{eqnarray}
C_{1p} &=& 0.03601  -0.00681 \Delta S + 0.00477\,\Delta T 
	+ \left(  \begin{array}{r}
	0 \\ 0.557 \\ 0.557
	\end{array} \right) \,\bar{\xi},   \\
C_{1n} &=& -0.49376 - 0.00366\,{\it \Delta T} + \left( \begin{array}{r}
	-0.835 \\  0.278 \\  0.278
	\end{array} \right) \,{\it \bar{\xi}}, 
\end{eqnarray}
\end{subequations}
and that for the weak charge of the cesium atom,  
$^{133}_{55}Cs$~\cite{noecker,wood}, is 
\begin{equation}
Q^{\rm SM}_{W}(^{133}_{55}Cs) =  -73.07 -0.749\,{\it \Delta S}
	- 0.046\,{\it \Delta T} 
	+ \left( \begin{array}{r}
	-130 \\ 105 \\ 105
\end{array} \right) \,{\it \bar{\xi}}.
\end{equation}
\subsubsection{Neutrino-quark scattering} 
For the neutrino-quark scattering, the experimental results
are given in terms of the model-independent parameters 
$g^2_{\alpha}$ and $\delta_{\alpha}^2$ $(\alpha =L,R)$~\cite{fh}, 
or their linear combination $K^{}_{\rm CCFR}$~\cite{ccfr}.
The definition and the theoretical prediction for 
$K_{\rm CCFR}$~\cite{ccfr} are
\begin{subequations}
\begin{eqnarray}
K_{\rm CCFR} &=&  1.7897 g_L^2 + 1.1479 g_R^2 - 0.0916 \delta_L^2 
	- 0.0782 \delta_R^2 \\
	&=& 0.5849          
	-0.0036\,{\it \Delta S} +0.0108\,{\it \Delta T} 
	+ \left( \begin{array}{r}
	-0.11 \\ -0.17 \\ -0.01
	\end{array} \right)\,{\it \bar{\xi}}.
\end{eqnarray}
\end{subequations}

Likewise, we find that the following combination $K^{}_{\rm FH}$ is 
most stringently constrained by the data of ref.~\cite{fh};
\begin{subequations}
\begin{eqnarray}
K_{\rm FH} &=& g_L^2 + 0.879 g_R^2 -0.010 \delta_L^2 -0.043 \delta_R^2 \\
	&=& 0.3309
	- 0.0018\,{\it \Delta S} +0.0060\,{\it \Delta T} 
	+ \left( \begin{array}{r}
	-0.13 \\ -0.09  \\ -0.05
	\end{array} \right)\,{\it \bar{\xi}}.
\end{eqnarray}
\end{subequations}
In the above, the predictions are evaluated 
at $\langle Q^2 \rangle_{\rm CCFR} = 35$ GeV${}^2$ 
and $\langle Q^2 \rangle_{\rm FH} = 20$ GeV${}^2$, respectively.
\section{Results}
\clean
It is now straight forward to obtain the constraints on the
parameters $\bar{\xi}$ and $T_{\rm new}$ from the electroweak
data listed in Table~\ref{tab:lep_table}. In the following analysis 
we use the direct constraints on the SM parameters
\begin{subequations}
\begin{eqnarray}
m^{}_t &=& 175.6 \pm 5.5{\rm \;GeV}~\,\cite{cdfd0}, \\
1/\bar{\alpha}(m^2_{Z_1}) &=& 128.75 
\pm 0.09\makebox[6.1mm]{}~\cite{pdg96}, \\
\alpha_s(m^{}_{Z_1}) &=& 0.118 \pm 0.003\makebox[6.1mm]{}~\cite{ej95},
\end{eqnarray}
\label{eq:direct}
\end{subequations}
$\!\!$and parametrize the $m^{}_H$-dependence of the results by using
the parameter $x^{}_H$~(\ref{eq:xh}). The allowed $m^{}_H$ range,
$m^{}_H>77\;$GeV~\cite{mh77} from the LEP 
search and $m^{}_H\:$\lsim $\:$ 150 GeV from the 
perturbative GUT condition~\cite{kane}, corresponds to 
\begin{equation}
-0.26<x^{}_H \;\lsim\; 0.41.
\label{eq:regxh}
\end{equation}
In subsection 4.1, we present the SM fit as a reference.
In subsection 4.2, we study the three leptophobic $Z'$ models
A, B and C, and in subsection 4.3, we find the lower mass bounds
for the heavier mass eigenstate, $Z_2$.
\subsection{SM}
The five-parameter fit for the $Z$ boson parameters and the direct
constraints~(\ref{eq:direct}) gives 
\begin{subequations}
\begin{eqnarray}
& & \left. 
\begin{array}{cclcl}
S_{\rm new} &=& -0.08-0.09x^{}_H & \pm & 0.14 \\ 
T_{\rm new} &=& -0.13+0.08x^{}_H & \pm & 0.16 
\end{array}
\right\}
\rho_{\rm corr} = 0.70, \\ & &  
\chi^2_{\rm min}/ ({\rm d.o.f.}) = (15.9-0.2x^{}_H)/(11).
\end{eqnarray}
\end{subequations}
The four-parameter fit for the LENC data and the direct
constraints on $m^{}_t$ and $1/\bar{\alpha}(m^2_{Z_1})$ gives
\begin{subequations}
\begin{eqnarray}
& & \left. 
\begin{array}{cclcl}
S_{\rm new} &=& -1.53-0.09 x^{}_H  & \pm & 1.16 \\ 
T_{\rm new} &=& -0.98+0.08 x^{}_H  & \pm & 0.54 
\end{array}
\right\}
\rho_{\rm corr} = 0.71, \\ & &  
\chi^2_{\rm min}/ ({\rm d.o.f.}) =  (1.9+0.0x^{}_H)/(5).
\end{eqnarray}
\end{subequations}
The $m^{}_W$  gives the constraint
\begin{equation}
-0.288S_{\rm new} + 0.418T_{\rm new} + 0.337 U_{\rm new} 
= 0.024+0.062x^{}_H \pm 0.091.
\end{equation}
The six-parameter fit to all the data of Table~\ref{tab:lep_table} 
and the constraint~(\ref{eq:direct}) gives
\begin{subequations}
\begin{eqnarray}
& & \left.\hspace{-15mm}
\begin{array}{r c r c l }
S_{\rm new} &=&  -0.12-0.09x^{}_H & \pm &  0.14 \\ 
T_{\rm new} &=&  -0.19+0.07x^{}_H & \pm &  0.15 \\
U_{\rm new} &=&  \makebox[3.3mm]{ }0.24+0.01x^{}_H & \pm & 0.27
\end{array}
\right\} \rho_{\rm corr}  = \left(
\begin{array}{rrr}
1.00 &  0.68 & -0.12  \\
     &  1.00 & -0.26  \\
     &       &  1.00   \end{array} \right) , \\
& & \chi^2_{\rm min}/ ({\rm d.o.f.})  = (20.0-0.2x^{}_H) / (18).
\end{eqnarray}
\end{subequations}
When $S_{\rm new}=U_{\rm new}$= 0, we find 
\begin{equation}
T_{\rm new} = -0.07+0.14x^{}_H \pm 0.11, 
\;\; \chi^2_{\rm min}/ ({\rm d.o.f.})=(21.4+0.9x^{}_H)/ (20).
\label{eq:sm_tnew}
\end{equation}
For the SM case, $S_{\rm new} = T_{\rm new} = U_{\rm new}$= 0, 
we find that the three-parameter fit gives 
$\chi^2_{\rm min}/({\rm d.o.f.})= 21.8/(21)$ for $x^{}_H$ = 0. 
This result is given in Table~\ref{tab:lep_table}.
\subsection{\bf $Z'$ models}
In this subsection, we show the constraints on the three $Z'$ models.
In the following, we set $S_{\rm new} = U_{\rm new}$ = 0, 
and allow the five-parameters, $m^{}_t$, $\alpha_s(m^{}_{Z_1})$, 
$\bar{\alpha}(m^2_{Z_1})$, $\bar{\xi}$ and $T_{\rm new}$ to vary
freely to fit the electroweak data of Table~\ref{tab:lep_table}
and the direct constraints~(\ref{eq:direct}), for each model at 
a fixed $m^{}_H$. The best-fit results at $m^{}_H = 100\;$GeV 
($x^{}_H$ = 0) under the constraint 
$T_{\rm new} \geq 0 $~(\ref{eq:TSMb}) are
shown in Table~\ref{tab:lep_table}.
\subsubsection{\bf Model A~\cite{eta_model}}
The five-parameter fit result can be parametrized as
\begin{subequations}
\begin{eqnarray}
& & \left. 
\begin{array}{cclcl}
T_{\rm new} &=& -0.073\;\: +0.142x^{}_H & \pm & 0.110 \\ 
\bar{\xi}   &=& \makebox[3.3mm]{ } 0.0016
	+0.0003x^{}_H & \pm & 0.0028 
\end{array} \right\}
\rho_{\rm corr} = 0.02, \\ & & 
\chi^2_{\rm min}/ ({\rm d.o.f.}) = (21.0+0.8x^{}_H) /(19).
\end{eqnarray}
\end{subequations}
The result for $x^{}_H$ = 0 is shown in Fig.~\ref{fig:cont}a.
Both $T_{\rm new}$ and $\bar{\xi}$ are consistent with
zero in the range~(\ref{eq:regxh}).
The $\chi^2_{\rm min}$ does not improves from the SM since
the theoretical prediction of $Q^{}_W(^{133}_{55}C_s)$ are worsening the 
fit; see Table~\ref{tab:lep_table}.
No other noticeable improvements can be observed in the pull factors
of Table~\ref{tab:lep_table}. 
When we impose the condition $T_{\rm new}\geq 0$~(\ref{eq:TSMb}), 
the $\chi^2_{\rm min}$ takes a minimum value, 21.4,
which is almost independent of $x^{}_H$ 
in the range $0.05\!<\!x^{}_H$ \lsim 0.41. 
Because d.o.f. of the 
fit decreases by two whereas $\chi^2_{\rm min}$ decreases only
by unity, the probability of the fit is worse than that of the SM.
\begin{figure}[btp]
\begin{center}
$\begin{array}{cc}
\leavevmode\psfig{figure=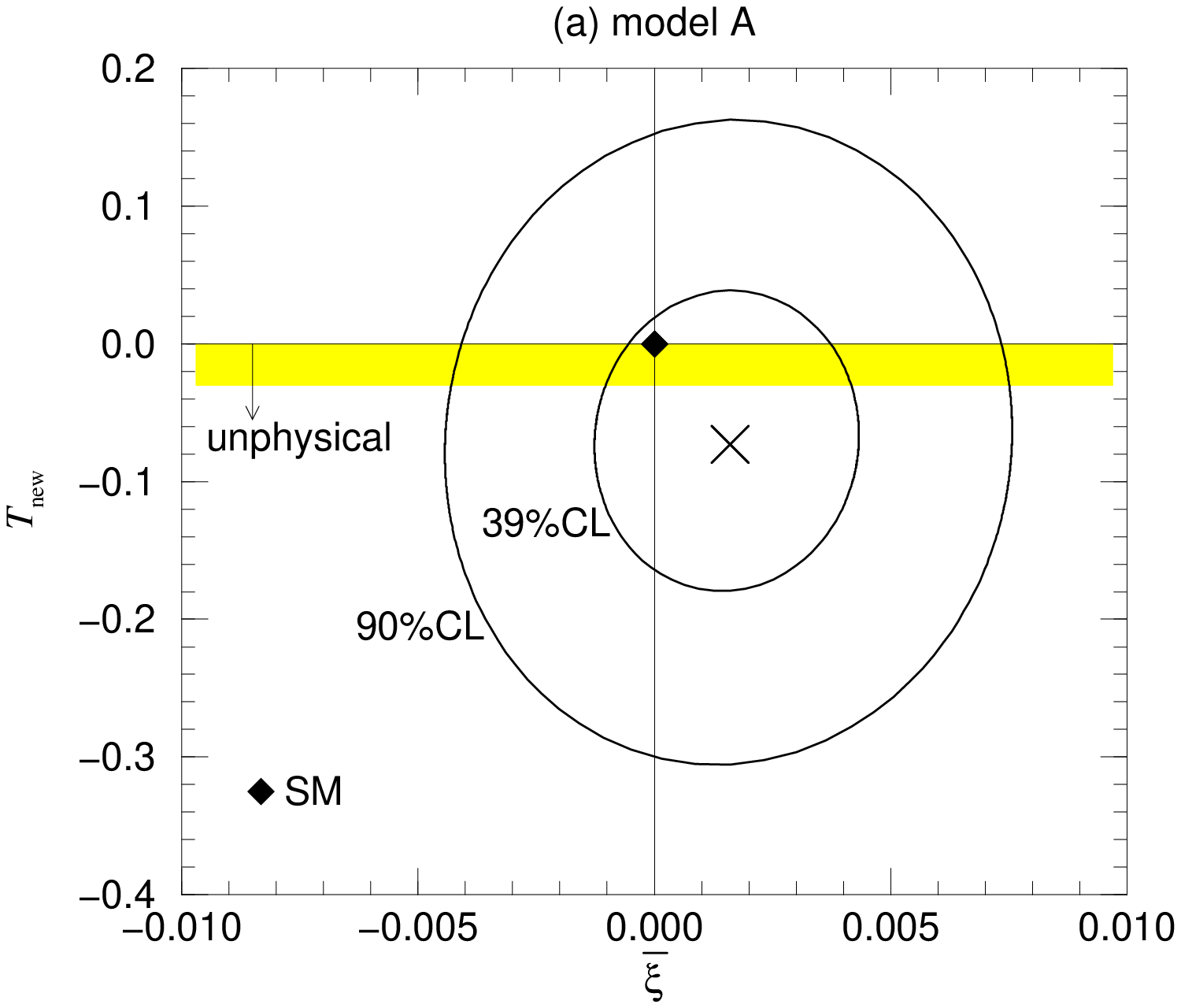,width=7cm} &
\leavevmode\psfig{figure=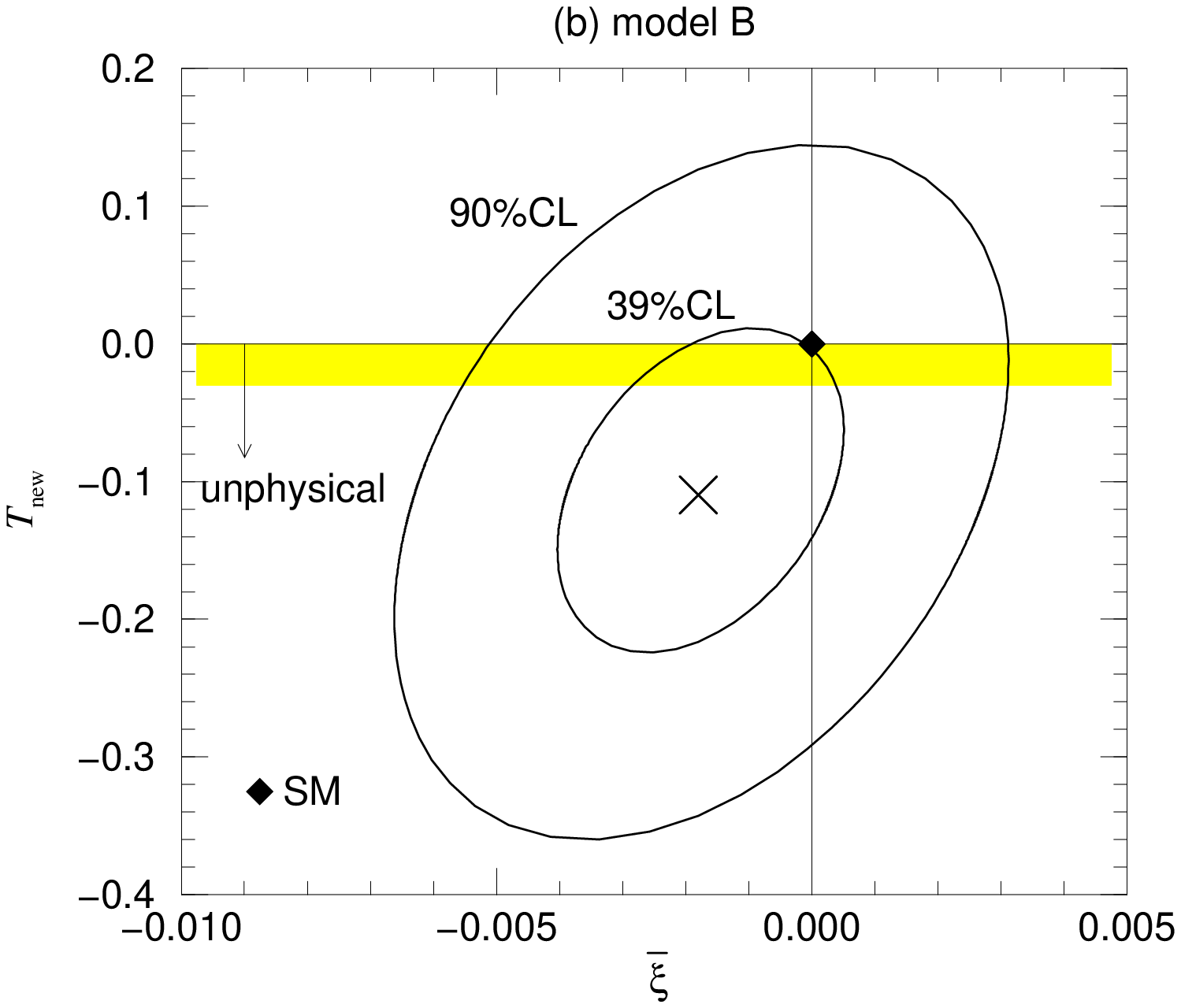,width=7cm}
\end{array}$
\leavevmode\psfig{figure=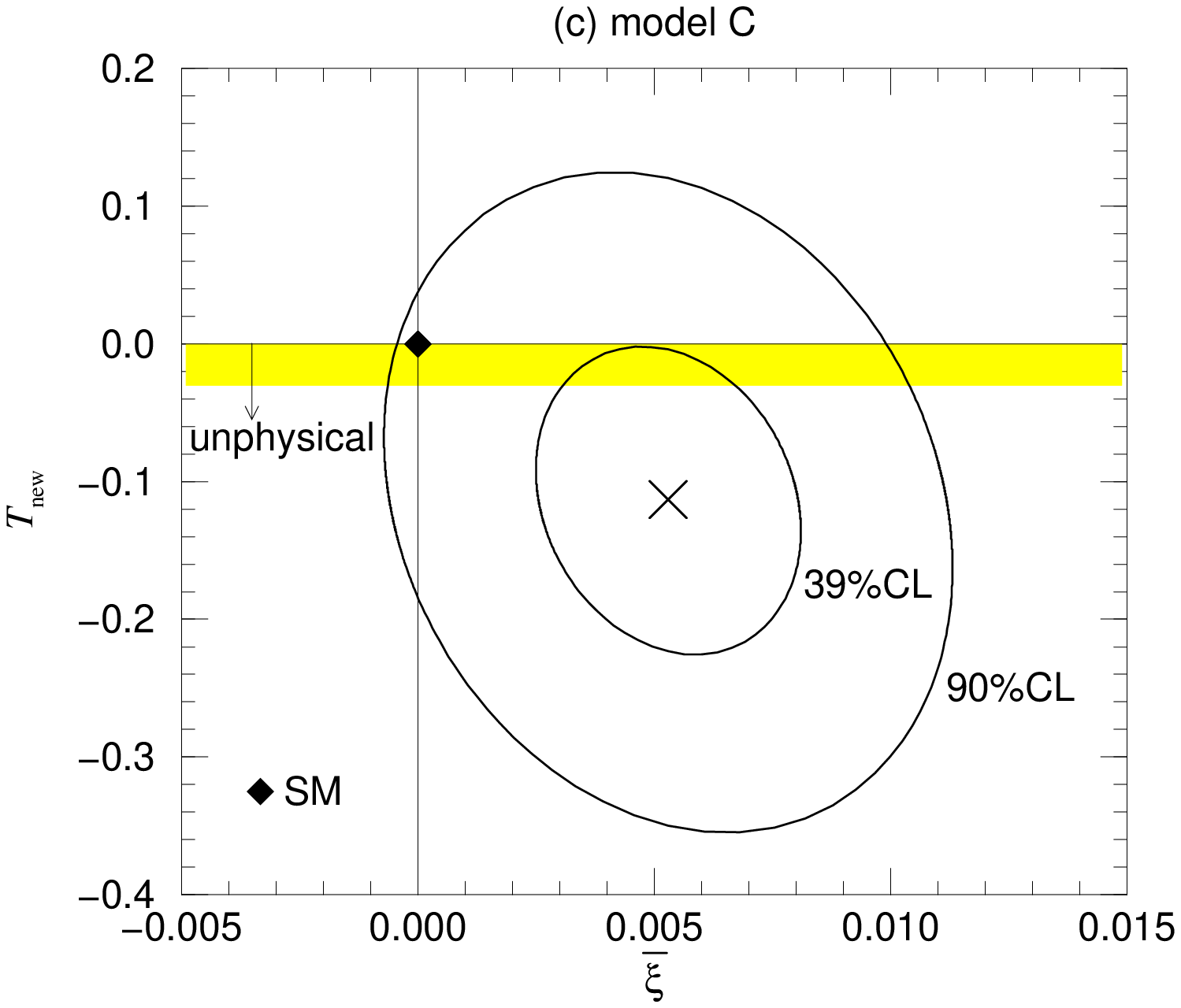,width=7cm}
\end{center}
\caption{{\small The $\bar{\xi}$ and $T_{\rm new}$ fit to  
all electroweak data. In all cases $m_t^{}$=175.6 $\pm$ 5.5 GeV, 
$\alpha_s^{}(m^{}_{Z_1})$ = 0.118 $\pm$ 0.03 and 
$1/\bar{\alpha}(m^2_{Z_1})$ = 128.75 $\pm$ 0.09 are taken 
as the external constraints. Also we assume 
$S_{\rm new} = U_{\rm new} = 0 $ 
and $m^{}_H = 100$ GeV.
The region of $T_{\rm new}< 0$ is unphysical in each case. 
The inner and outer contours correspond to 
$\Delta \chi^2$ = 1 ($\sim$ 39$\%$ CL) and $\Delta \chi^2$ = 4.61 
($\sim$ 90$\%$ CL), respectively. The minimum of $\chi^2$ is marked 
by the sign '$\times$', whose magnitudes are 21.0, 20.7 and 17.9
for the models A, B and C, respectively.
The point $\bar{\xi}=T_{\rm new}=0$ represents 
the SM result ($\chi^2_{\rm min} \sim$ 21.8).}} 
\label{fig:cont}
\vspace{10mm}
\end{figure}
\subsubsection{\bf Model B~\cite{flipped}}
The five-parameter fit gives
\begin{subequations}
\begin{eqnarray}
& & \left. 
\begin{array}{cclcl}
T_{\rm new}  &=& -0.110\;\:+0.154x^{}_H & \pm & 0.119 \\ 
\bar{\xi}    &=& -0.0018+0.0006x^{}_H & \pm & 0.0023 
\end{array}
\right\}
\rho_{\rm corr} = 0.38, \\ & &  
\chi^2_{\rm min}/ ({\rm d.o.f.}) = (20.7+1.3x^{}_H) /(19).
\end{eqnarray}
\end{subequations}
The result for $x^{}_H$ = 0 is shown in Fig.~\ref{fig:cont}b.
In the model B, not only $T_{\rm new}$ and $\bar{\xi}$ are consistent
with zero but also the $\chi^2_{\rm min}$ does not improve
from its SM value~(\ref{eq:sm_tnew}).
When we impose the condition $T_{\rm new}\geq 0$, 
the $\chi^2_{\rm min}$ takes a minimum value, 21.4, 
which is almost independent of $x^{}_H$   
in the range $0.19\!<\!x^{}_H\lsim 0.41$. No noticeable improvement nor
worsening of the fit is found from the pull factors listed in
Table~\ref{tab:lep_table}.
\subsubsection{\bf Model C~\cite{lbl}}
\begin{figure}[t]
\begin{center}
$ \begin{array}{cc}
\leavevmode\psfig{figure=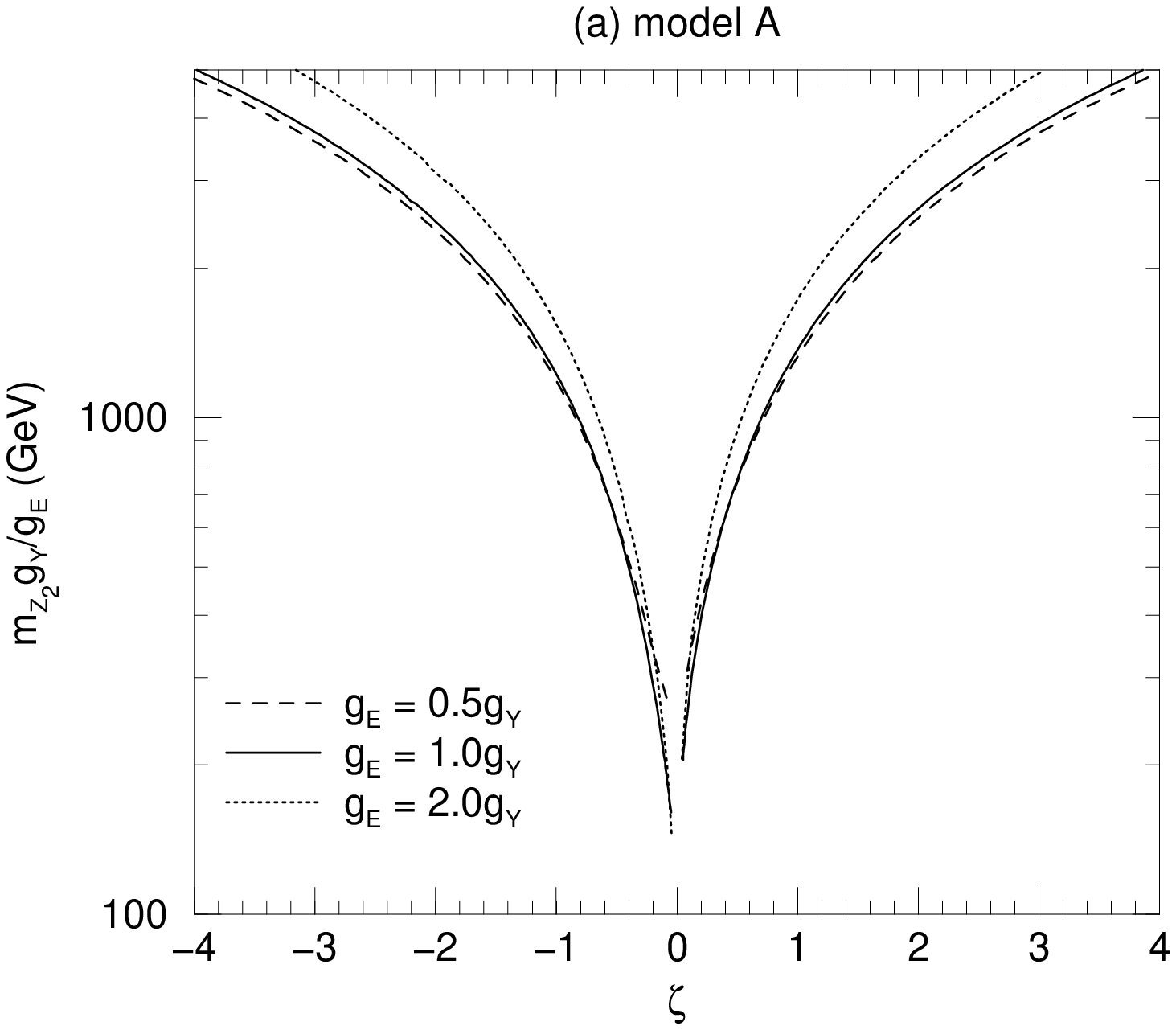,width=7cm} &
\leavevmode\psfig{figure=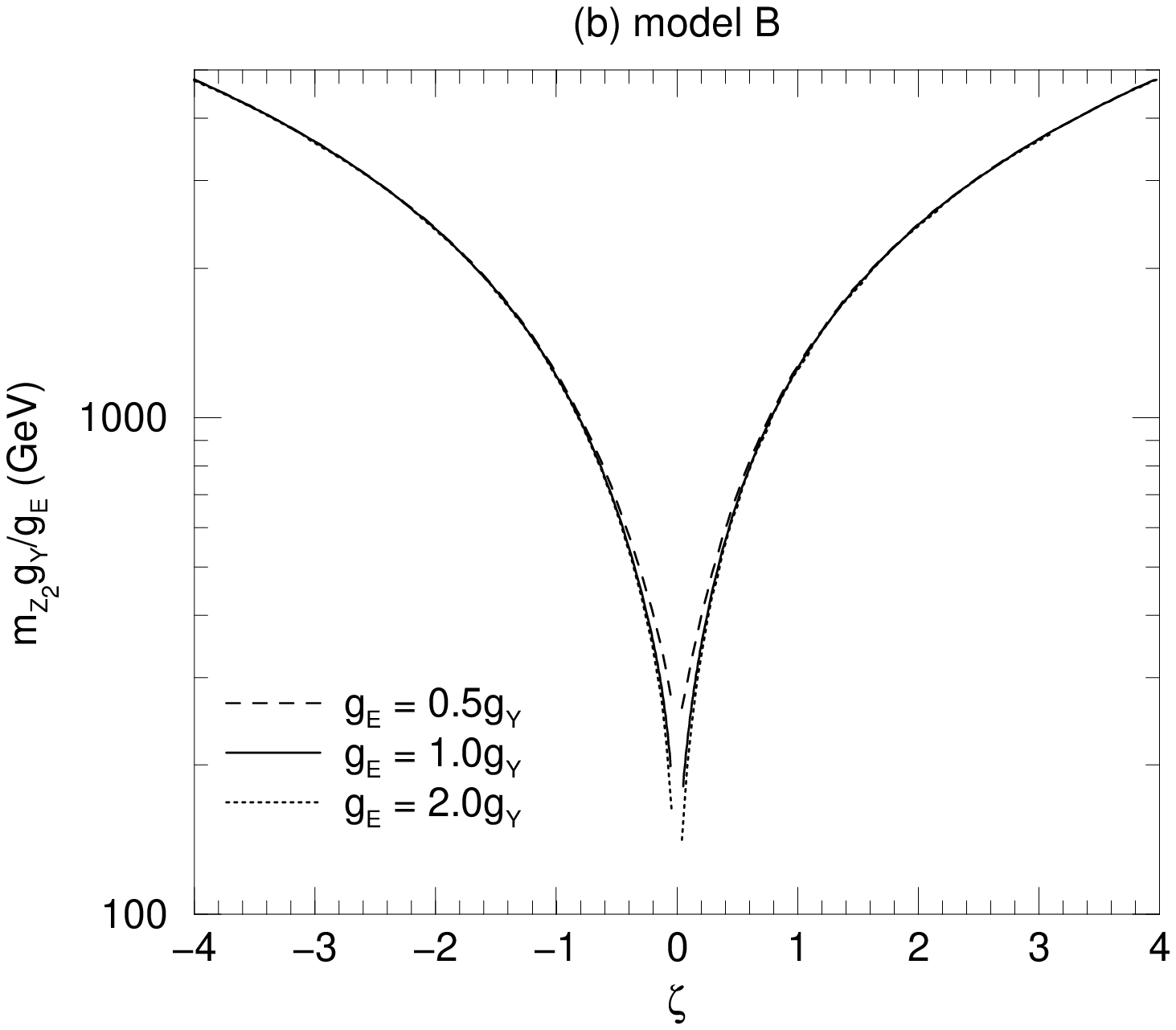,width=7cm} \\
\leavevmode\psfig{figure=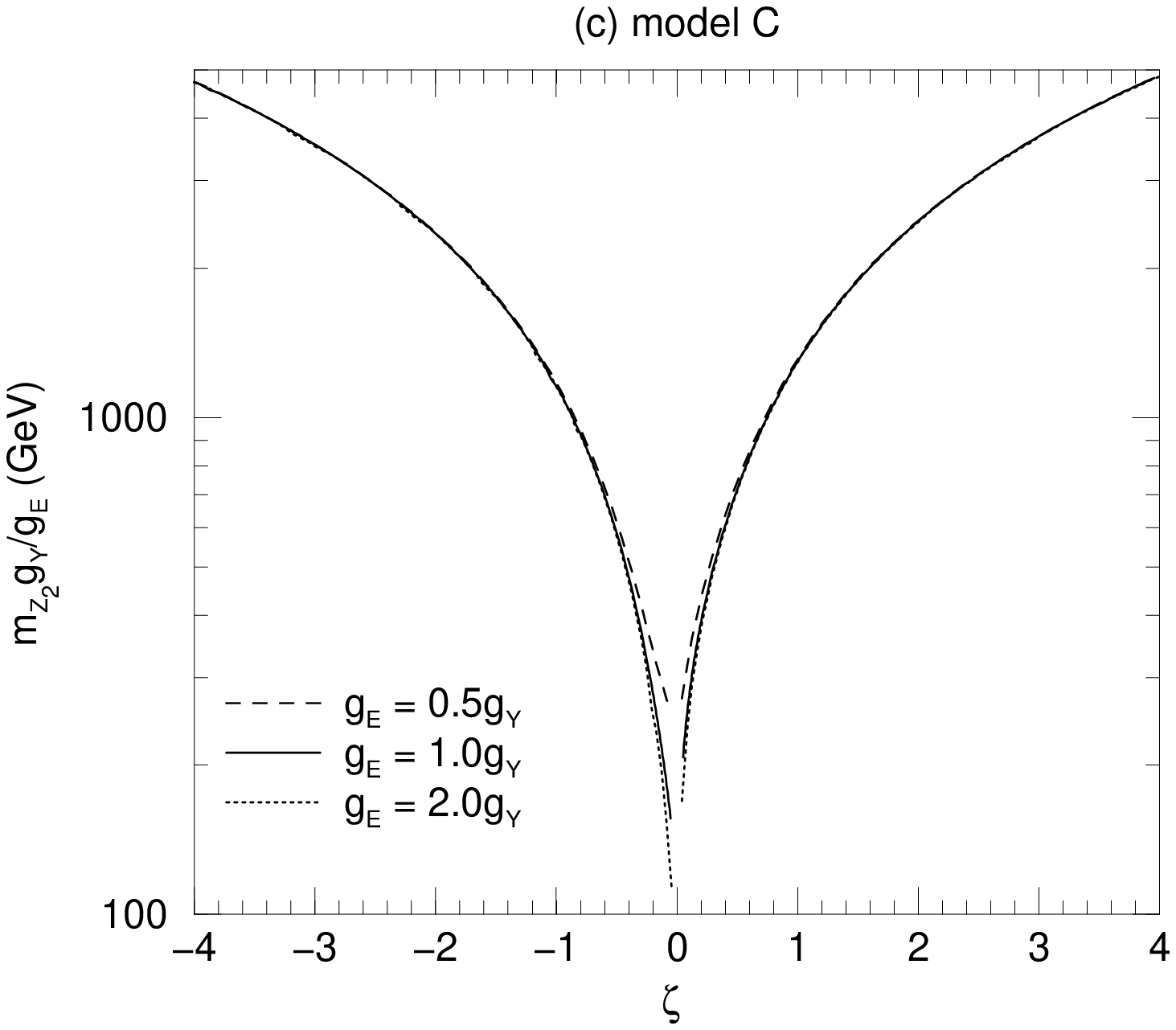,width=7cm} &
\leavevmode\psfig{figure=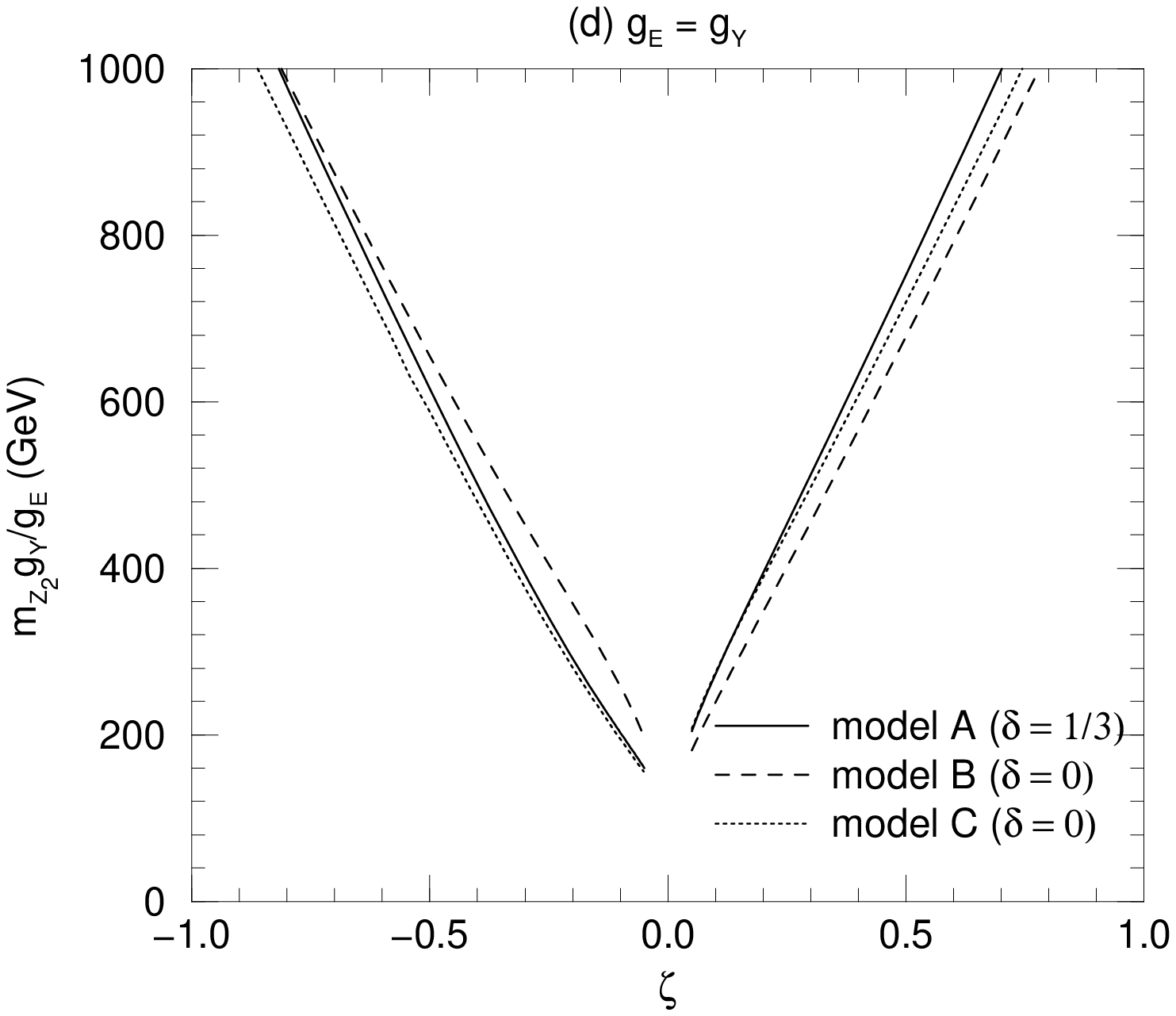,width=7cm} 
\end{array} $
\end{center}
\caption{{\small The 95$\%$ CL lower mass limit
of the heavier mass eigenstate
$Z_2$ as a function of the effective mixing parameter 
$\zeta$ in~(\ref{eq:zeta}).
}}  
\label{fig:Z2}
\vspace{10mm}
\end{figure}
The five-parameter fit gives
\begin{subequations}
\begin{eqnarray}
& & \left.
\begin{array}{cclcl}
T_{\rm new} &=& -0.112\;\:+0.143x^{}_H & \pm & 0.112 \\ 
\bar{\xi}   &=&  \makebox[3.3mm]{ } 0.0052
-0.0001x^{}_H  & \pm & 0.0028
\end{array}
\right\}
\rho_{\rm corr} = -0.18, \\ & & 
\chi^2_{\rm min}/ ({\rm d.o.f.}) = (17.9+1.1x^{}_H) /(19).
\end{eqnarray}
\end{subequations}
The result is shown in Fig.~\ref{fig:cont}c for $x^{}_H$ = 0
($m^{}_H$ = 100 GeV).
In the model C, the positive value of the mixing parameter
$\bar{\xi}$ is favored ($\sim 2 \sigma$), and $\chi^2_{\rm min}$
improves by about three from that of the SM value~(\ref{eq:sm_tnew}).
From the pull factors listed in Table~\ref{tab:lep_table},
we find that the improvement of the fit occurs for 
two observables, $R_b$ and $Q^{}_W(^{133}_{55}C_s)$, 
as originally arranged in ref.~\cite{lbl}.
When we impose the condition $T_{\rm new}\geq 0$, 
the $\chi^2_{\rm min}$ takes a minimum value, 18.6, 
which is almost independent of $x^{}_H$ 
in the range $0.20\!<\!x^{}_H$ \lsim 0.41.
The probability of the fit, however, does not improve 
significantly over that of the SM.
\subsection{\bf $Z_2$ mass bounds}
The above study shows that no significant improvement over the
SM is found for the three leptophobic $Z'$ models in the
latest electroweak data.
In this subsection, we obtain the 95$\%$ CL lower mass 
limit of the heavier mass eigenstate $Z_2^{}$.
From eq.~(\ref{eq:approx_Txi}), it is clear that no constraint 
is found at $\zeta = 0$. At a fixed non-zero $\zeta$,
the  parameter $T_{\rm new}$ and $\bar{\xi}$, which parametrize 
the effect of the $Z$-$Z'$ mixing, 
can be expressed in terms of the ratio of the physical masses 
$m^2_{Z_1}/m^2_{Z_2}$ for a given value of $g^{}_E/g^{}_Y$.
For each $\zeta$ and $g^{}_E/g^{}_Y$, we can express 
the $\chi^2$-function in terms of the positive parameter
$m^2_{Z_1}/m^2_{Z_2}$. 
We then find the 95$\%$ CL upper bound $r^{}_{95}$ on the 
ratio $r = m^2_{Z_1}/m^2_{Z_2}$ from the condition 
\begin{equation}
\int^{\infty}_{r_{95}}dr e^{-\frac{\chi^2(r)}{2}} = 
0.95\int^{\infty}_{0}dr e^{-\frac{\chi^2(r)}{2}}.
\end{equation}
The 95$\%$ lower mass bound for $Z_2$ is then obtained as 
$m^{}_{Z_1}/\sqrt{r^{}_{95}}$. The results are shown 
in Fig.~\ref{fig:Z2}a, b and c for the models
A, B and C, respectively. As may be expected from the 
small mixing formulae~(\ref{eq:approx_Txi}), the approximate scaling
low is found for the $g^{}_E/g^{}_Y$-dependence of the limit:
$m^{}_{Z_2}g^{}_Y/g^{}_E$ is roughly independent 
of $g^{}_E/g^{}_Y$. In Fig.~\ref{fig:Z2}d, we show the 
small $\zeta$ region more clearly by using the linear 
scale for the three models at $g^{}_E = g^{}_Y$.
The $Z'$ boson with $m^{}_{Z_2} < 1$ TeV is allowed by the 
electroweak data (for $g^{}_E = g^{}_Y$) only when 
the effective mixing parameter $\zeta$~(\ref{eq:zeta})
satisfies the following conditions:
$-0.8\!<\!\zeta\!<\!0.7$, $-0.8\!<\!\zeta\!<\!0.8$ 
and $-0.9\!<\!\zeta\!<\!0.8$ 
for the models A, B and C, respectively. 

Throughout our analysis, we have neglected the effects of the
direct exchange of the $Z_2$ boson, which is proportional to 
the mixing angle $\xi$. We find that the 95$\%$ CL lower bounds
on $m^{}_{Z_2}$ are slightly weakened by taking account of 
such effects, at most 3 GeV for $|\zeta |$ = 0.05.
For larger $|\zeta |$, the effect is negligible because of 
higher lower mass bound of $m^{}_{Z_2}$.

The lower mass limit of $Z_2$ also depends on the Higgs boson mass.
Because of the condition $T_{\rm new} \ge 0$~(\ref{eq:TSMb}) and
large $m^{}_H$ makes the best-fit value of $T_{\rm new}$
large, 
the $Z_2$ mass bound tends to weak
as the Higgs boson mass increases.
At $|\zeta|$ = 1, the mass bounds for $m^{}_H$ = 80~(150) GeV 
are  $4\%$~($8\%$) severer~(weaker) than that for $m^{}_H$ = 100 GeV.
\section{Summary}

 In this paper, we have investigated the constraints on the 
three leptophobic $Z'$ models from the latest electroweak data.
The $Z'$ boson in the model A~\cite{eta_model} is 
essentially $Z_{\eta}$ 
of the string-inspired $E_6$ model
with large kinetic mixing, that in the model B~\cite{flipped} 
couples only to the decouplet of the 
flipped SU(5)$\times$U(1) GUT. In the model C~\cite{lbl}
the $Z'$ couplings to quarks are determined by refering to 
the electroweak data of 1995.
In our parametrization, the $Z$-$Z'$ mixing effects 
are parametrized in terms of the effective mass mixing parameter
$\bar{\xi}$ and the non-SM contribution $T_{\rm new}$
to the electroweak $T$ parameter due to the mass
shift $\Delta m^2 = m^2_{Z_1}-m^2_{Z}$.
Since the mass shift $\Delta m^2$
is negative,  $T_{\rm new}$ is positive definite.
Compact parametrizations of the predictions of the SM and
the three leptophobic $Z'$ models are given for all the 
electroweak observables.
From the fit to the latest electroweak data, we find that 
none of the three models gives a significantly improved 
fit over the SM. The improvement in $\chi^2_{\rm min}$ is
found to be at most three (for the model C) while each model has 
two additional parameters, $\bar{\xi}$ and $T_{\rm new}$.

Finally, we have obtained the 95$\%$ CL lower mass 
limit of the heavier mass eigenstate $Z_2^{}$.
When the mixing parameter $\zeta$~(\ref{eq:zeta}) is large,
$|\zeta |>1$, the lower mass bound exceeds 1 TeV for all the 
models. The leptophobic $Z'$ boson lighter than 1 TeV is allowed
only in the range
$-0.8\!<\!\zeta\!<\!0.7$, $-0.8\!<\!\zeta\!<\!0.8$ and 
$-0.9\!<\!\zeta\!<\!0.8$ 
for the models A, B and C respectively,
for $g^{}_E = g^{}_Y$ and $m^{}_H$ = 100 GeV.

\section*{Acknowledgment}
We would like to thank Seiji Matsumoto for useful comments.
The work of G.C.C. was supported in part by Grant-in-Aid for Scientific 
Research from the Ministry of Education, Science and Culture of Japan.


\end{document}